\documentclass{amsart}

\usepackage{amsmath}
\usepackage{amsfonts}

\usepackage{natbib}
\usepackage{subcaption}
\usepackage{graphicx}
\usepackage{epsfig}
\newtheorem{theorem}{Theorem}
\newtheorem{lemma}{Lemma}

\DeclareMathOperator*{\argmax}{arg\,max}

\title{An easy-to-use empirical likelihood ABC method}

\author[Chaudhuri]{Sanjay Chaudhuri}
\address{Department of Statistics and Applied Probability, National University of Singapore, Singapore 117546}
\email{sanjay@stat.nus.edu.sg}
\author[Ghosh]{Subhroshekhar Ghosh}
\address{Department of Mathematics, National University of Singapore, Singapore 117546.}
\email{subhrowork@gmail.com}
\author[Nott]{David J. Nott}
\address{Department of Statistics and Applied Probability, National University of Singapore, Singapore 117546}
\email{standj@nus.edu.sg}
\author[Pham]{Kim Cuc Pham}
\address{Department of Statistics and Applied Probability, National University of Singapore, Singapore 117546}
\email{staptkc@nus.edu.sg}

\begin{document}

\begin{abstract}
Many scientifically well-motivated statistical models in natural, engineering and environmental sciences are specified through a generative process, but in some cases it may not be possible to
write down a likelihood for these models analytically.  Approximate Bayesian computation (ABC) methods, which allow Bayesian inference in these situations, are 
typically computationally intensive.  Recently, computationally attractive empirical likelihood based ABC methods have been suggested in the literature.
These methods heavily rely on the availability of a set of suitable analytically tractable estimating equations. 
We propose an easy-to-use empirical likelihood ABC method, where the only inputs required are a choice of summary statistic, it's observed value, and the ability to simulate summary
statistics for any parameter value under the model.  It is shown that the posterior obtained using the proposed method is consistent, and its performance is explored using various examples.
\end{abstract}





\maketitle





The concept of likelihood is central to parametric statistical inference. However, for many models encountered in natural, 
engineering and environmental sciences, it is difficult to express the likelihood analytically.  These models are often specified in a generative fashion, 
so that independent samples can be generated from these models for any value of the model parameters.
Approximate Bayesian computation (ABC) methods are useful for Bayesian inference in situations
like these
\citep{tavare+bgd97,beaumont+zb02,marin+prr11,fearnhead+p12,blum+nps13}.  
Simple ABC approaches involve first simulating parameter values and data from the prior, and then reducing the data 
to a lower-dimensional summary statistic which is informative for the parameter.
Following this, a comparison is made between
simulated and observed summary statistics.  For 
simulated summary statistics sufficiently close to the observed value, the corresponding parameter value
is accepted as an approximate draw from the posterior.  Other generated values of the parameters are discarded.   
This basic rejection ABC algorithm can be cast as a special case of importance sampling for
a kernel approximation of a summary statistic likelihood, and there is a known curse of dimensionality
associated with use of such methods.  
More sophisticated sampling algorithms somewhat improve efficiency 
\citep{marjoram+mpt03,sisson+ft07,beaumont+rmc09}, but even state-of-the-art ABC methods are computationally demanding 
in high-dimensional cases.  

Partly in response to the above difficulties, various pseudo-likelihood based methods have been considered.  Several such likelihoods have already been used for non-generative models by various authors \citep{monahan+b92,lazar03,chaudhuri+g11}.  Many of these approaches can also be employed in cases where a generative model exists but the associated likelihood is intractable.

Among the pseudo-likelihood methods used for generative models, perhaps the most popular is the synthetic likelihood introduced by \citet{wood10}, which uses a working multivariate normal model for the summary statistics.  Its 
Bayesian implementation is discussed in detail in \citet{price+dln16}.
The synthetic likelihood sometimes performs poorly when the normal approximation of the distribution of the summary statistics is inaccurate.  \citet{wood10} explores marginal transformation of the summaries to make the normality assumption more reasonable. However, such marginal transforms cannot usually achieve multivariate normality when the dependence structure is non-normal, or guarantee validity of the normal approximation over the whole parameter space.  
Extensions that relax the requirement of normality have been a continuous topic of interest for many researchers in this area.
\citet{Fasiolo2016} consider an extended saddlepoint approximation, whereas, \citet{Dutta2016} propose
a method based on logistic regression.  
  Using auxiliary parametric models, \citet{drovandi+pl15} describe an encompassing framework for many of the above suggestions, which they call parametric Bayesian indirect inference.

A fast empirical likelihood based ABC approach was recently suggested by
\citet{mengersen+pr13}, where the intractable likelihood for the generative process
was replaced by an appropriate non-parametric empirical likelihood. 
Empirical likelihood \citep{owen01} is computed from a constrained estimator of the joint empirical distribution function of the data. 
By using this likelihood \citet{mengersen+pr13} could avoid any assumption of normality of the summary statistics.  However, in their proposal constraints based on analytically tractable estimating functions of both the data and the parameters were required.  Since such functions are not readily available, their proposed method is not always easy to apply. 

In this article, we introduce an easy-to-use empirical likelihood based ABC method, where the only required inputs are a choice of summary statistic, it's observed value, 
and the ability to simulate that particular statistic under the model for any parameter value.  
Although we refer to our method as an empirical likelihood ABC approach, it differs from the classical ABC algorithms, in the sense
that no kernel approximation of the summary statistic likelihood is involved.  Furthermore, unlike \citet{mengersen+pr13}, the proposed method does not require analytically tractable estimating functions involving the parameters. That is, the proposed method is an interpretable likelihood-based, completely data dependent ABC procedure.

The detailed implementation of our proposal is inspired by some algorithms for computation of marginal maximum likelihood or MAP estimates and the Fisher information in
complex latent variable models \citep{doucet+gr02,lele+dl07}.  \citet{lele+dl07} have used the name ``data cloning'' for this idea.  
In the case of finding a marginal maximum likelihood estimate, for example, we can consider an experiment where several copies of the 
data are to be observed, and then we suppose that all the copies turned out to be equal to the observed data.  As the number of copies increases, then the likelihood based on the imaginary
replicates concentrates on the global modes of the likelihood for the original problem.  Furthermore, we may be able to compute other quantities such as the Fisher information
by the device of considering the imaginary replicates.   The key idea in all these approaches is that some features of the likelihood of interest can be related to the 
likelihood for an artificial problem where some imaginary replications of the data are supposed to have occurred.  We consider in this paper something roughly similar to obtain empirical likelihood constraints which involve sums of terms
for independent data replicates, and where the terms can be re-weighted in the usual empirical likelihood
fashion.  The likelihood for this artificial problem has features that are related to the original likelihood, but how to construct an empirical likelihood analogue in the artificial problem with replication 
is clear, whereas this is usually not the case for the original problem of interest.  

In the next section we describe the basic intuition of the approach, and Section 3 gives the definition of our proposed empirical likelihood approximation.  Section 4 discusses the choice of estimating equations, and Section 5 describes basic asymptotic properties of the method, proving posterior consistency under reasonable conditions.  Section 6 considers four examples and Section 7 gives some concluding discussion.

\section{An easy-to-use Bayesian empirical likelihood method}\label{sec:multsamp}

In this section we explain the basic idea of the proposed method.  This involves considering an artificial experiment incorporating some data replicates.  
  The likelihood for the replicates can then be related to the original observed data likelihood.  

We consider a set of $n-$dimensional random vectors $\left\{X_i(\theta), i\in\mathbb{N}_o,\theta\in\Theta\right\}$, where $\mathbb{N}_o=\{o\}\cup\mathbb{N}$, i.e. the set of positive integers appended with symbol $o$, and for every $\theta$, $\left\{X_i(\theta), i\in\mathbb{N}_{o}\right\}$ are i.i.d. with density $f_{\theta}$.  
Suppose $Y_i=Y_i(\theta_o)$, $i=1,2,\ldots,m$ are i.i.d. $f_{\theta_o}$, where $\theta_o$ is the true value of the parameter.  For a chosen $\theta\in\Theta$, let $X_i=X_i(\theta)$, $i=1,\dots,m$ be the specific values simulated i.i.d. as $f_{\theta}$.  The vectors $X_i$ and $Y_j$ are independent, for any $i,j\in\{1,2,\ldots,m\}$.   
That is, by construction, the likelihood of $\theta$ based on each pair $(Y_i,X_i)$, $i=1,2,\dots,m$, turns out to be:
$$f_{\theta}^{\otimes2}\left(Y_i,X_i\right)=f_{\theta}\left(Y_i\right)f_{\theta}\left(X_i\right).$$
Now the likelihood based on all $m$ pairs when each $Y_i$ is observed to be $X_o$ is given by:
\begin{equation}
l^{(m)}(\theta)=\prod^m_{i=1}f_{\theta}^{\otimes2}\left(X_o,X_i\right)=\left\{f_{\theta}\left(X_o\right)\right\}^m\prod^m_{i=1}f_{\theta}\left(X_i\right). \label{eq:mlik}
\end{equation}
We focus on the corresponding scaled log-likelihood:
\begin{equation}
\frac{1}{m}\log\left\{l^{(m)}(\theta)\right\}=\log\{f_{\theta}\left(X_o\right)\}+\frac{1}{m}\sum^m_{i=1}\log\left\{f_{\theta}\left(X_i\right)\right\}. \label{eq:lmlik}
\end{equation}
As $m\rightarrow\infty$ the second term in the right hand side of \eqref{eq:lmlik} converges to the differential entropy $E_{\theta}[\log f_\theta(X)]=\log C(\theta)$ (say), 
and we notice that for large $m$ the right hand side of \eqref{eq:lmlik} doesn't depend on the particular values
$X_1,\dots,X_m$ generated.  Furthermore, in the limit the second term $\log C(\theta)$ is in general a slowly changing function of $\theta$ compared to $\log f_\theta(X_o)$. 
  In particular, for a location model $C(\theta)$ is a constant independent of $\theta$.  Thus, for location models, with $\pi(\theta)$ as the preassigned prior supported on $\Theta$, upon normalisation, the term $C(\theta)$ cancels out in the limit of the corresponding posterior, 
\[
\lim_{m\rightarrow\infty}\frac{\exp\left(\frac{1}{m}l^{(m)}(\theta)\right)\pi(\theta)}{\int_{t\in\Theta}\exp\left(\frac{1}{m}l^{(m)}(t)\right)\pi(t)dt},
\]
making it equal to the posterior conditional only on the observed data.  This suggests that, in general, we can use 
  $\log l^{(m)}(\theta)/m$ as an approximation to the observed data log likelihood $\log f_\theta(X_o)$.
Our motivation for considering $\log l^{(m)}(\theta)$ as defined in \eqref{eq:lmlik} is that it can be easily estimated solely from observed $X_o$ and the generated $X_1$, $\ldots$, $X_m$ using the empirical likelihood based methods, which we describe next.



\section{Definition of the ABC empirical likelihood}

Suppose $F_\theta^{\otimes 2}$ is the distribution corresponding to $f_{\theta}^{\otimes2}$. When $\theta=\theta_o$, $(Y_i,X_i)$, $i=1,\dots, m$ 
are independent observations generated from $F_{\theta_o}^{\otimes 2}$.  
Based on these $m$ data points we estimate $F^{\otimes 2}_{\theta}$, with each $Y_i=X_o$ by an empirical likelihood obtained under judicious choices of constraints which depend only on $X_o$ and $X_1$, $\ldots$, $X_m$.

Suppose $g_1,\ldots,g_r$ are deterministic functions of the observations. 
By construction, when $\theta=\theta_o$, $X_o$, $X_1$, $\ldots$, $X_m$ are identically distributed.  Then for any $k=1,\ldots,r$, and $i=1,\dots, m$, 
\begin{equation}
  E\left[g_k\left(X_{i}(\theta_o)\right)-g_k\left(X_{o}(\theta_o)\right)\right]=0. \label{eq:ex}
\end{equation}
We base our constraints on these functions which play the role of the summary statistics for the data.  Let us define a $r$-dimensional vector valued function $g(x)=(g_1(x),\dots, g_r(x))^T$ and for each $i=1$, $2$, $\ldots$, $m$, 
\begin{equation*}
  h_i(\theta)=g\left(X_i(\theta)\right)-g\left(X_o(\theta_o)\right).
  \end{equation*}

For any $\theta\in\Theta$, define the random set:
\begin{align}
~&\mathcal{W}_{\theta}=\left\{w~:~\sum^m_{i=1}w_ih_i(\theta)=0\right\} \cap\Delta_{m-1}\label{eq:w1}\\
=&\bigcap^r_{k=1}\left\{w~:~\sum^m_{i=1}w_i\left[g_k\left(X_{i}(\theta)\right)-g_k\left(X_{o}(\theta_o)\right)\right]=0\right\} \cap\Delta_{m-1},\nonumber 
\end{align}  
where $\Delta_{m-1}$ is the $m-1$ dimensional simplex.  

Based on observations $(X_o,X_1)$, $\ldots$, $(X_o,X_m)$, the  distribution $F^{\otimes 2}_{\theta}$ is estimated by the empirical distribution constrained by the set $\mathcal{W}_{\theta}$.  This estimate puts weight $\hat{w}_i$ on points $(X_o,X_i)$ for each $i=1,\ldots,m$, where the vector of weights $\hat{w}$ is constrained to be in $\mathcal{W}_{\theta}$.  The optimal weights $\hat{w}$ are given by 
\begin{align}
  \hat{w}:=\hat{w}(\theta):=&\hat{w}(X_1(\theta),\ldots,X_m(\theta),X_o(\theta_o))=\argmax_{w\in\mathcal{W}_{\theta}}\left(\prod^m_{i=1}mw_i\right). \label{eq:w2}
\end{align}
If the problem in \eqref{eq:w2} is infeasible, $\hat{w}$ is defined to be zero.

Once $\hat{w}$ is determined, the left-hand side of \eqref{eq:lmlik} is estimated by:
\begin{equation*}
\widehat{\frac{1}{m}\log\left(l^{(m)}(\theta)\right)}=\frac{1}{m}\sum^m_{i=1}\log(\hat{w}_i(\theta)).
\end{equation*}
Now, in conjunction with the prior $\pi(\theta)$, we can define a posterior $\Pi(\theta\mid X_o)$ of the form,
\begin{align}
\Pi(\theta\mid X_o):=&\frac{\left[e^{\frac{1}{m}\sum^m_{i=1}\log\left(\hat{w}_i(\theta)\right)}\right]\pi(\theta)}{\int_{t\in\Theta}\left[e^{\frac{1}{m}\sum^m_{i=1}\log\left(\hat{w}_i(t)\right)}\right]\pi(t)dt}\propto\left[e^{\frac{1}{m}\sum^m_{i=1}\log(\hat{w}_i)}\right]\pi(\theta).
\label{eq:mpost}
\end{align}
When $\prod^m_{i=1}\hat{w}_i=0$, we define $\Pi(\theta\mid X_o)=0$. 
 
Inference about the true value of the parameter can be drawn from the posterior $\Pi(\theta\mid X_o)$. Clearly, since each $\hat{w}_i$ is bounded, the estimated likelihood is bounded for all values of $\theta$.  Thus the posterior $\Pi(\theta\mid X_o)$ is proper for any proper prior $\pi$.  
No analytic expression for this posterior exists in general.  However, using Markov chain Monte Carlo (MCMC) techniques, a sample of any required size can be drawn from $\Pi(\theta\mid X_o)$, which is sufficient for making posterior inferences. 
All components of $\hat{w}$ in \eqref{eq:w2} are strictly positive iff the origin is in the interior of the convex hull defined by the vectors $h_1$, $h_2$, $\ldots$, $h_m$.  
When the origin is at the boundary of this convex hull, the constrained optimisation in \eqref{eq:w2} is still feasible, but some of the estimated weights are zero, so by our definition the posterior is zero as well.  In both these cases, $\mathcal{W}_{\theta}$ in \eqref{eq:w1} is non-empty.  If the origin is outside this closed convex hull, 
this optimisation problem is infeasible and again by definition the value of the posterior is zero.

Even though the proposed method is similar in spirit to the synthetic likelihood, it is more general than the latter.  
Synthetic likelihood assumes normality of the joint distribution of the summary statistics.  
Even though many summary statistics are asymptotically normally distributed, this is not always the case, and  
in some cases involving non-normal summary statistics the synthetic likelihood can perform poorly.  
\citet{mengersen+pr13} use Bayesian empirical likelihood in an ABC setting. However, the estimating equations they use directly depend on the parameter, 
and these equations must be analytically specified.  Such estimating equations may not be available in many problems.  
In our empirical likelihood approximation, we only require the observed data $X_o$ and simulated data $X_1,\dots,X_m$ under the model for a given $\theta$.  
Furthermore, the proposed empirical likelihood can be computed quite easily and usually at a reasonable computational cost.
The proposed empirical likelihood estimates joint weights by matching the moments of $g(X_1)$, $\ldots$, $g(X_m)$ with that of $g(X_o)$, without requiring a direct relationship with the parameter.
Finally, the proposed likelihood in \eqref{eq:mpost} is different from the original empirical likelihood defined in \citet{owen01} and thus would differ from the latter in both asymptotic and finite sample properties.


\section{Choice of Estimating Equations}
It is clear that much depends on the correct specification of the constraints imposed on the weights which determine the empirical likelihood. 
In most applications of Bayesian empirical likelihood, these constraints directly depend on the parameter $\theta$ through an analytically specified estimating equation.  
However, the structure of our proposed empirical likelihood allows us to specify constraints without involving the parameter except through the simulation of the observations $X_i=X_i(\theta)$.
Many choices for the  constraint functions are possible.  We outline some simple choices below. 

From now on, we assume that for $i\in\{o,1,\dots,m\}$, $X_i\in\mathbb{R}^n$.  For some $k$ and some positive deterministic $\gamma_k$, for each $i$ we may define,
\begin{equation}
g_k\left(X_{i}\right)=\frac{1}{n}\sum^n_{j=1}X^{\gamma_k}_{ij}, \label{eq:mnt}
\end{equation}
so that $g_k$ is the $\gamma_k$th raw sample moment.  Provided $E[X_{ij}^{\gamma_k}]$ exists, such a choice of $g_k$ would constrain the underlying distribution through its moments.  
Similarly the $\gamma_k$ sample quantile of $X_i$ may be used for any $\gamma_k \in[0,1]$, which would directly put a constraint on the distribution through its quantiles.
Another possibility is the proportion of times $X_i$ is larger than $\gamma_k$, 
\begin{equation}
g_k\left(X_{i}\right)=\frac{1}{n}\sum^n_{j=1}\mathbf{1}_{\{X_{ij}\ge \gamma_k\}}. \label{eq:upt}
\end{equation}
Other than these generic choices, one can base the constraints on functionals of transformed variables. For example, in certain situations constraints based on the spectral distribution of the data could be used. 

With these choices of $g$, the likelihood is estimated by matching the marginal moments, quantiles and up-crossings of the generated vectors with those of the observed values.  
In complex data models, where the $X_i$ and $X_o$ have non-identically distributed and dependent components,
looking at simple marginal properties of the components of $X_i$ and $X_o$ may not be adequate and some insight about the model could be used to choose the constraints.  In such cases, constraints 
can be based on joint moments, joint quantiles or joint up-crossings of subsets of $\{X_{i1},\ldots,X_{in}\}$, as we illustrate later.  
Any summary statistics used in traditional ABC analyses can also be used in the proposed empirical likelihood approach. 

\section{Asymptotic Properties}
The asymptotic properties of conventional ABC methods have been a topic of much recent research
\citep{frazier+mrr18,li+f18a,li+f18b}.  Here 
we investigate some basic asymptotic properties of our proposed empirical likelihood method.  
Following \citet{owen01} the weights in \eqref{eq:w2} can be obtained by maximising the objective function:

\[
L(w)=\sum^m_{i=1}\log(mw_i)-\alpha\left(\sum^m_{i=1}w_i-1\right)-n\lambda^T\sum^m_{i=1}w_ih_i,
\]

where $\alpha$ and $\lambda$ are the Lagrange multipliers associated with the constraints.  

It is easily shown that $\alpha=1$ and the optimum weights are given by:
\[
\hat{w}_i=\frac{1}{m}\cdot\frac{1}{1+\hat{\lambda}^Th_i},
\]
where $\hat{\lambda}$ is obtained by solving the equation:
\[
\sum^m_{i=1}\frac{h_i}{1+\hat{\lambda}^Th_i}=0.
\]

In what follows below, we consider limits as $n$ and $m=m(n)$ grow unbounded.  Furthermore, for convenience, we make the dependencies of $X_o$ and $X_1$, $X_2$, $\ldots$, $X_m\in\mathbb{R}^n$ on sample size $n$ and parameter $\theta$ explicit. 
In what follows, a sequence of events $\{E_n, n\ge 1\}$ is said to occur with high probability, if $P(E_n)\rightarrow 1$ as $n\rightarrow\infty$.

Suppose that we define
\[
h^{(n)}_i\left(\theta\right)=\left\{g\left(X^{(n)}_i(\theta)\right)-g\left(X^{(n)}_o(\theta_o)\right)\right\}, 
\]  
and assume $E[g(X^{(n)}_i(\theta))]$ is finite so that we can write 
\[
g\left(X^{(n)}_i(\theta)\right)=E\left[g\left(X^{(n)}_i(\theta)\right)\right]+\xi^{(n)}_i(\theta)=\mathfrak{g}^{(n)}(\theta)+\xi^{(n)}_i(\theta),
\] 
where $E[\xi^{(n)}_i(\theta)]=0$ for all $i$, $n$ and $\theta$.
We make the following assumptions.
\begin{itemize}
\item[(A1)] (Indentifiability and convergence) There is a sequence of positive increasing real numbers $b_n\rightarrow\infty$, such that:
\[
\mathfrak{g}^{(n)}(\theta)=b_n\left\{\mathfrak{g}(\theta)+o(1)\right\},
\]   
where $\mathfrak{g}(\theta)$ is a one to one function of $\theta$ that does not depend on $n$.  Furthermore, $\mathfrak{g}(\theta)$ is continuous at $\theta_o$ and for each $\epsilon>0$,  and for all $\theta\in\Theta$, there exists $\delta>0$, such that whenever $\mid\mid\theta-\theta_o\mid\mid>\epsilon$, $\mid\mid \mathfrak{g}(\theta)-\mathfrak{g}(\theta_o)\mid\mid>\delta$.

\item[(A2)] (Feasibility) For each $\theta$, $n$ and $i=o,1$, $\ldots$, $m(n)$, the vectors $\xi^{(n)}_i(\theta)$ are identically distributed, supported over the whole space, and their distribution puts positive mass on every orthant, $\mathcal{O}_s$ of $\mathbb{R}^r$, $s=1$, $2$, $\ldots$, $2^r$.  Furthermore, for every orthant $\mathcal{O}_s$, as $n\rightarrow\infty$, 
\[
\sup_{\{i~:~\xi^{(n)}_i(\theta)\in\mathcal{O}_s\}}\mid\mid \xi^{(n)}_i(\theta)\mid\mid\longrightarrow\infty
\]
in probability, uniformly in $\theta$.   
\item[(A3)] (Growth of extrema of Errors) As $n\rightarrow\infty$, 
\[
\sup_{i\in\{o,1,2,\ldots, m(n)\}}\frac{\mid\mid \xi^{(n)}_i(\theta)\mid\mid}{b_n}\rightarrow 0
\]
in probability, uniformly in $\theta\in\Theta$.
\end{itemize}
Assumption (A1) ensures identifiability and additionally implies that $\mathfrak{g}^{(n)}(\theta)/b_n-\mathfrak{g}(\theta)$ converges to zero uniformly in $\theta$.
Assumption (A2) is important for ensuring that with high probability the empirical likelihood ABC posterior is a valid
probability measure for $n$ large enough.  Assumptions (A2) and (A3) also link the number of simulations $m$ to $n$ 
and ensure concentration of the posterior with increasing $n$.  
The proofs of the results below are given in the Appendix.  The main result, Theorem 1, shows posterior consistency for the proposed empirical likelihood method.  

Let $l_n(\theta):=\exp(\sum^{m(n)}_{i=1}\log\left(\hat{w}_i(\theta)\right)/m(n))$ and for each $n$, we define:
\[
\Theta_n=\left\{\theta~:~\mid\mid\mathfrak{g}(\theta)-\mathfrak{g}(\theta_o)\mid\mid\le b^{-1}_n\right\}.
\]
By continuity of $\mathfrak{g}$ at $\theta_0$, $\Theta_n$ is nonempty for each $n$.  Furthermore, since $b_n$ is increasing in $n$, $\Theta_n$ is a decreasing sequence of sets in $n$. 
\begin{lemma}\label{lem:1}
  Under assumptions (A1) to (A3), with high probability, the likelihood $l_n(\theta)>0$ for all $\theta\in\Theta_n$.
\end{lemma}

Lemma \ref{lem:1} shows that for large $n$ the estimated likelihood is strictly positive in a neighbourhood of $\theta_0$.  Next, we show that the empirical likelihood is zero outside certain neighbourhood of $\theta_0$.

For $\theta\in \Theta$ and $\epsilon>0$, by $B(\theta,\epsilon)$ we denote the ball of radius $\epsilon$ around $\theta$.
\begin{lemma}\label{lem:2}
  Under assumptions (A1) - (A3), for every $\epsilon>0$, the empirical likelihood is zero outside $B(\theta_0,\epsilon)$, with high probability.
\end{lemma}

Now suppose we choose $\epsilon=b^{-1}_1$ and $n>n(b^{-1}_1)$ 
such that $l_n(\theta)$ is positive on $\Theta_n$ with high probability. This proves that for large values of $n$, with high probability:
\[
\int_{\theta\in\Theta}l_n(\theta)\pi(\theta)d\theta\ge\int_{\theta\in\Theta_n}l_n(\theta)\pi(\theta)d\theta>0,
\]
and 
\[
\Pi_n\left(\theta\mid X_o(\theta_o)\right)=\frac{l_n(\theta)\pi(\theta)}{\int_{t\in\Theta}l_n(t)\pi(t)dt}
\]
is a valid probability measure (with high probability).  The main result, Theorem 1 below, establishes posterior consistency.

\begin{theorem}\label{thm:1}
As $n\rightarrow\infty$, $\Pi_n\left(\theta\mid X_o(\theta_o)\right)$ converges in probability to $\delta_{\theta_o}$, where $\delta_{\theta_0}$ is the degenerate probability measure supported at $\theta_0$. 
\end{theorem}

\section{Illustrative Examples and Applications}

In this section we consider four illustrative examples.  First, however, we comment on computational issues arising in their implementation.
The estimated weights in \eqref{eq:w2}, which define the empirical likelihood, can only be computed numerically in almost all cases. This makes it necessary to use methods such as MCMC to sample from the posterior.  
The support of the posterior may be non-convex \citep{chaudhuri+my17}.  
In the examples below, we use Metropolis-Hastings random walk methods with normal proposal 
for the MCMC sampling, but more sophisticated methods could also be used in the case of a high-dimensional parameter.   

The MCMC sampling procedure using the proposed empirical likelihood in effect samples from a likelihood estimated using
  Monte Carlo methods.
Similar to the Bayesian synthetic likelihood approaches \citep{price+dln16}, it can be thought of as an implementation of the pseudo-marginal Metropolis-Hastings method for a modified target distribution \citep{beaumont03,andrieu+r09}.  The number of replicates generated, i.e. $m$ should be chosen judiciously.
Even though empirical evidence suggests that the results are not statistically sensitive to the number of samples, the choice of $m$ has computational implications.  Several authors \citep{price+dln16,doucet+pdk15} have noted that a large variance of the noisy likelihood estimate results in a poorly mixing  MCMC chain.
The choices for $m$ used in the examples below are sufficient to ensure adequate mixing, but they
depend on the dimensionality and distributional properties of the summary statistics, and need to be considered on a case by case basis.



Computation of the empirical likelihood is generally very fast.  Several efficient optimisation methods are available.  We have used the R package {\tt emplik} \citep{zhou+y16} 
in the experiments below.  The computational
effort involved in implementing the proposed approach is similar to the synthetic likelihood in our examples.

Four examples are considered.  The first is a simple normal location example, and we use this to illustrate the effects of different
summary statistic choices in the method.  The second example concerns a $g$-and-$k$ model, which is a standard
benchmark model for ABC inference algorithms.  The third example is a dependent data example, for an ARCH(1) model - 
this was also considered in \cite{mengersen+pr13}.  The summary statistics used in this example are non-Gaussian, and
we show that synthetic likelihood does not work well here, but the empirical likelihood is more robust to the non-normality.  
The fourth example is a real example for stereological extremes.  We use this example for two purposes. First of all, we find summaries for which the proposed method performs comparably to the synthetic likelihood and rejection ABC methods.  Furthermore, in order to illustrate
the importance of the choice of the summary statistics, we consider a set of hard to match summaries, which fit poorly to the assumed model. It is seen that the proposed empirical likelihood does not work well in this situation. However, it is no worse than the synthetic likelihood.  It is difficult to implement the latter with the same summary statistics as well.


\begin{table}[t]
\def~{\hphantom{0}}
\caption{The coverage and the average length of $95\%$ credible intervals for $\mu$ for various choices of constraint functions when $\mu=0$ and $n=100$.  The coverage for the true posterior is $0.95$ and average length is $0.39$ (2 d.p.).}{%
\begin{tabular}{lcc}
Constraint Functions & Coverage & Average Length\\
Mean, (a).&$0.93$&$0.34$\\
Median, (e).&$0.93$&$0.43$\\
First two raw moments, (a), (b).&$0.88$&$0.30$\\
First three raw moments, (a), (b), (c).&$0.85$&$0.27$\\
Three quartiles, (e), (f), (g).&$0.76$&$0.28$\\
Mean and Median, (a), (e).&$0.76$&$0.24$\\
First four raw moments, (a), (b), (c), (d).&$0.72$&$0.22$\\
\end{tabular}
}
\label{Tab2}
\end{table}

\subsection{Normal distribution}

Our first example considers inference about a mean $\mu$ for a random sample of size $n=100$ from a normal density, $N(\mu,1)$.  
The prior for $\mu$ is $N(0,1)$.  The observed data $X_o$ is generated with $\mu=0$. The exact posterior for $\mu$ is  
normal, $N(\sum^n_{j=1}X_{oj}/(n+1),(n+1)^{-1})$.  The proposed empirical likelihood based method was implemented with $m=25$.  We considered several choices of constraint functions  $g_1$, $\ldots$, $g_r$.  
Specifically, for $i=o,1,\ldots,m$, we take (a) $g_1(X_i)=\sum^n_{j=1}X_{ij}/n$,  (b) $g_2(X_i)=\sum^n_{j=1}X^2_{ij}/n$, (c) $g_3(X_i)=\sum^n_{j=1}X^3_{ij}/n$, (d) $g_4(X_i)=\sum^n_{j=1}X^4_{ij}/n$, (e) $g_5(X_i)=\mbox{median of }X_i$, (f) $g_6(X_i)=\mbox{first quartile of }X_i$, (g) $g_7(X_i)=\mbox{third quartile of }X_i$.  Here the constrains considered use the
first four raw moments ((a)-(d)) and the three quartiles ((e)-(g)).
Combinations of these constraints are considered within the empirical likelihood procedure.  
  
Different constraints are compared based on the coverage and the average length of the $95\%$ credible intervals for $\mu$ obtained from $100$ replicates.  These values give some indication of frequentist coverage of the credible
intervals when $\mu=0$, but the results can also be used to compare with corresponding quantities for the true posterior as one way of checking if the empirical likelihood approach approximates the true posterior well in relevant ways for inference.
For each replicate, MCMC approximations to the posterior are
based on $50,000$ sampling iterations with $50,000$ iterations burn in.  
The results are given in Table~\ref{Tab2}.

From Table \ref{Tab2}, we see that the proposed method performs quite well when either the mean or median is used as constraint function.  Note that the sample mean is minimal sufficient for $\mu$, and would be an ideal choice of summary statistic
in conventional likelihood-free procedures such as ABC.  Table \ref{Tab2} also shows that when 
many summary statistics are used, the performance of empirical likelihood ABC deteriorates.    
Inclusion of raw moments of higher orders and more quantiles makes both frequentist performance (in terms of coverage) and any correspondence with the true posterior worse.  Simultaneous constraints with the mean and median gives a coverage and average credible interval length quite different to those for the true posterior.  This is consistent with the experiences of \citet{mengersen+pr13}, who implement a Bayesian empirical likelihood based on parametric constraints.  

  Unlike the synthetic likelihood, which can automatically down-weight relatively uninformative summaries through the estimation of their means and covariances, the empirical likelihood based method, as proposed, cannot choose constraints and therefore is more vulnerable to uninformative components.
On the other hand, the empirical likelihood does not assume normality for summary statistics, and performs better in models where normality should not be assumed, (see example C. below).
For the proposed empirical
likelihood method, similar to conventional ABC methods, we recommend to use summary statistics that are informative and
of minimal dimension.
Finally, we note that increasing the value of $m$ beyond $25$ seemed to cause no appreciable difference in the results.



\begin{figure}
  \begin{center}
\resizebox{3.25in}{3.25in}{\includegraphics{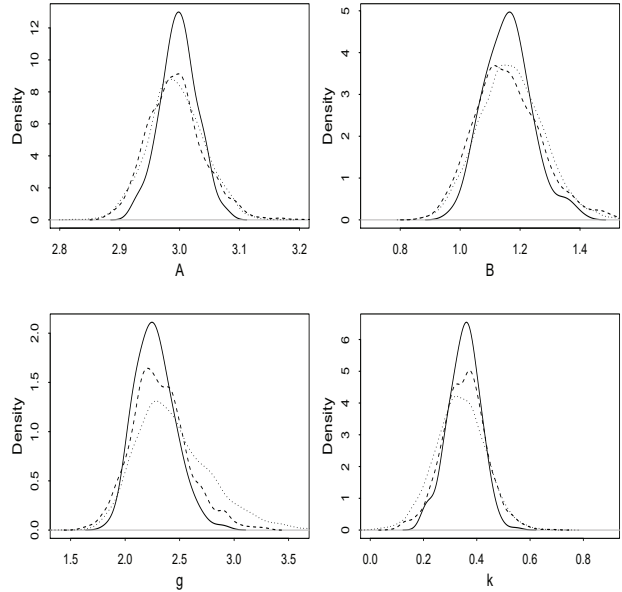}}
\caption{Estimated marginal posterior densities by proposed method (solid), synthetic likelihood (dashed) and regression ABC (dotted) for parameters of the $g$-and-$k$ model.}
\label{F2}
\end{center}
\end{figure}

\subsection{$g$-and-$k$ distribution}
Our second example concerns inference for the $g$-and-$k$ distribution \citep{haynes1997robustness}.  This
distribution is defined through its quantile function,

\begin{align}
~&Q(p;A,B,g,k) =A+B \left[ 1+c\times\frac{1-\exp\left\{-gz(p)\right\}}{1+\exp\left\{-gz(p)\right\}} \right] \left\{ 1+{z(p)}^2 \right\} ^kz(p),\nonumber
\end{align} 
where $z(p)$ is the $p$th standard normal quantile and conventionally $c$ is fixed at $0.8$, which results in the constraint $k>-0.5$.
Simulation from this model can be performed by 
transforming uniform random variables on $[0,1]$ by the quantile function.  This feature, and the fact that 
there is no closed form expression for the density function, makes likelihood-free inference methods attractive.
Components of the parameter vector $\theta=(A,B,g,k)$ are respectively related to location, scale, skewness and kurtosis of the distribution. In the  ABC context, this distribution was first considered in \citet{allingham2009bayesian}, 
with an analysis of the related $g$-and-$h$ distribution given earlier in \citet{peters+s06}.

A dataset of size $n=1000$ was simulated from the distribution with $(A,B,g,k)=(3,1,2,0.5)$.  A uniform prior $U(0,10)^4$ for $\theta$ was assumed.  We approximate the proposed empirical likelihood and the synthetic likelihood using $m=40$ datasets each of length $n$ for each value of $\theta$.  The mean and the three quartiles were used as summary statistics.   Some summary statistics used in \citet{drovandi2011likelihood} based on octiles were also considered, but 
resulted in slightly inferior performance for estimation of the kurtosis parameter $k$.  
Posterior samples were drawn using a random walk Metropolis algorithm with normal proposal and diagonal
proposal covariance matrix, with the variances chosen based on a pilot run.  Posterior summaries are based on $100,000$ sampling
iterations after $100,000$ iterations burn in.  
 
The results are presented Figure~\ref{F2}.  Estimated marginal posterior densities obtained from the synthetic likelihood and proposed empirical likelihood are shown as dashed and solid lines respectively. Also shown is 
a ``gold standard''  answer based on rejection ABC with a small tolerance and linear
regression adjustment \citep{beaumont+zb02}.  For the ABC approach, to improve computational efficiency, we restricted the prior for $\theta$ from $U(0,10)^4$ to $U(2,4)\times U(0,2)\times U(0,4)\times U(0,1)$.  This restricted prior is broad enough to contain the support of the posterior based on the original prior.  The ABC estimated marginal posterior densities (dotted) shown in Figure~\ref{F2} were based on $5,000,000$ samples, choosing the tolerance so that $2000$ samples are kept. The summary statistics used are asymptotically normal and $n$ is large, so the synthetic likelihood is expected to work well in this example, which it does. Our proposed method gives comparable results to synthetic likelihood and
the ``gold standard'' ABC analysis, although there does seem to be some slight underestimation of posterior uncertainty in the empirical likelihood method, similar to the normal location example.



\begin{figure}[t]
  \begin{center}
\resizebox{3.25in}{3.25in}{\includegraphics{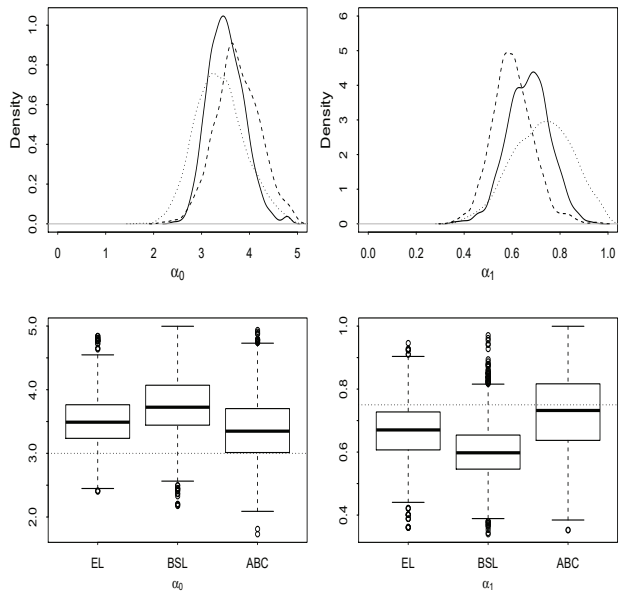}}
\caption{Estimated marginal posterior densities of parameters $\alpha_0$ and $\alpha_1$ in the ARCH(1) model.  The top row
shows kernel density estimates (empirical likelihood ABC (solid), synthetic likelihood (dashed), rejection ABC (dotted)), while the bottom row shows boxplots of posterior samples.  In the boxplots, the horizontal
dotted lines show the true parameter values.}
\label{F3}
\end{center}
\end{figure}

\subsection{An ARCH(1) model}
In contrast to the previous example, we now consider an example with summary statistics which are not close to normal, so that 
the assumptions behind the synthetic likelihood are not satisfied.  
We consider an autoregressive conditional heteroskedasticity or ARCH(1) model, where for each $i=o,1,2,\ldots,m$, the components $X_{i1},X_{i2},\ldots,X_{in}$ are dependent.  This model was also considered in \citet{mengersen+pr13}. 
For each $i$, the time series ${X_{ij}}_{1\leq j \leq n}$ is generated by 
\begin{equation}\label{eq:arch_mod}
X_{ij}=\sigma_{ij}\epsilon_{ij},  \;\; {\sigma_{ij}}^2=\alpha_0+\alpha_1{X_{i(j-1)}}^2.
\end{equation}
where the $\epsilon_{ij}$ are iid $N(0,1)$ random variables. Here $\alpha_0,\alpha_1>0$ and stationarity requires
$\alpha_1<1$.
To simulate $X_{ij}$, $j=1,\dots, n$ for each $i$, we first simulate $\epsilon_{ij}$, for $j=1,...,n$.  We set the initial standard deviation $\sigma_{i1}$ to $\surd\{\alpha_0/(1-\alpha_1)\}$, which is consistent
with stationarity \citep[Section 21]{hamilton1994time}, and then the data can be generated following \eqref{eq:arch_mod}.  The parameter vector $(\alpha_0,\alpha_1)$ is given a uniform prior over $(0,5)\times(0,1)$.

Our summary statistics are the lag 1 autocovariance of the squared data, together with the three quartiles 
of the absolute values of the data.   
The summary based on the autocovariance is needed here, since the data are dependent. The quartiles of the absolute values of the data provide some information about the marginal distribution.
Our observed data were of size $n=1000$, with $(\alpha_0,\alpha_1)=(3,0.75)$ and we used $m=20$ replicates for each
likelihood approximation for both empirical and synthetic likelihoods in Bayesian computations.  

Marginal posterior densities were estimated for the parameters based on $100,000$ sampling iterations with $100,000$ iterations burn in for both the synthetic likelihood and proposed empirical likelihood.   We compare these methods with the posterior obtained using rejection ABC with $1,000,000$ samples, a tolerance of $0.0025$ and linear regression adjustment.  
The estimated marginal densities in Figure \ref{F3} for the proposed method are quite close to the ABC gold standard.  The synthetic likelihood estimated marginal posterior densities are quite different to those obtained by ABC however, 
especially for $\alpha_1$.  In this example 
the first order autocorrelation statistic is highly non-Gaussian, so the normality assumption made in the synthetic likelihood formulation is not satisfied.  

\subsection{Stereological data}  

Next we consider an example concerning the modelling of diameters of inclusions (microscopic particles introduced
in the steel production process) in a block of steel.  The size of the largest
inclusion in a block is thought to be important for steel strength.  
The data considered here were first analysed by \citet{anderson2002largest}, and consist of measurements on inclusions
from planar cross-sections.  \citet{anderson2002largest} considered a spherical model for the inclusions, which
leads to a model with a tractable likelihood.  \citet{bortot2007inference} later extended this to an elliptical
inclusion model which does not have tractable likelihood, and it is this model that we discuss.

\citet{anderson2002largest} assume that the inclusion centres follow a homogeneous Poisson process with rate $\lambda$. 
In the elliptical model, for each inclusion the three principal diameters of the ellipse 
are assumed independent of each other and of the process of inclusion centres.  
Let $V$ be the largest inclusion diameter for a given inclusion.  Given $V$, the two other principal diameters are determined by multiplying $V$ with an independent uniform $U[0,1]$ random variable.
The diameter $V$, conditional on exceeding a threshold value $v_0$ ($5\mu m$ in \citet{bortot2007inference}) is assumed to follow a generalised Pareto distribution:
\begin{displaymath}
\operatorname{pr}(V\leq v|V>v_0)=1-\left\{1+\frac{\xi (v-v_0)}{\sigma}\right\} _{+}^{-\frac{1}{\xi}}.
\end{displaymath}
Since the inclusion centres follow a homogeneous Poisson process, so do the inclusions 
with $V>v_0$. The parameters of the model are given by $\theta=(\lambda,\sigma,\xi)$.
We assume independent uniform priors for $\lambda$, $\sigma$ and $\xi$ with ranges $(1,200)$, $(0,10)$ and $(-5,5)$ respectively. 
A detailed implementation of ABC for this example is discussed in \citet{erhardt+s15}.

\begin{figure}[t]
  \begin{center}
\resizebox{3in}{3in}{\includegraphics{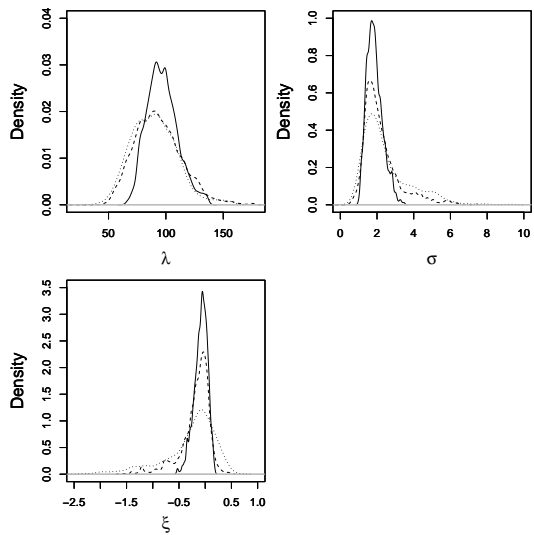}}
\caption{Estimated marginal posterior densities of $\lambda$, $\sigma$ and $\xi$ using empirical likelihood ABC (solid), rejection ABC (dotted) and synthetic likelihood (dashed).}
\label{density_inclu}
\end{center}
\end{figure}

The observed data consists of $n=112$ observations, measuring the largest principal diameters of elliptical cross-sections of inclusions for a planar slice. The number of inclusions in each dataset generated from the model is random.  Writing $L$ for the number of inclusions, the summary statistics used are $a)$ $(L-112)/100$, $b)$ the mean of the observed planar measurements, $c)$ the median of the of the observed planar measurements, and $d)$ the proportion of planar measurement less than or equal to six (approximately the median for the observed data). 
 Even though the number of observations is itself random, the estimating equations we use are unbiased under the truth.

Using the summary statistics described above, we compare the proposed empirical likelihood based method with the synthetic likelihood and a gold standard rejection ABC algorithm with small tolerance and linear regression adjustment. For the 
rejection ABC method we generated $10,000,000$ samples from the elliptic inclusion model and use a tolerance of $0.00005$ and linear regression adjustment.
Both the proposed empirical likelihood and the synthetic likelihood methods use $m=25$ samples.  
In total, $25,000$ samples were drawn from the empirical likelihood and synthetic likelihood posterior densities, following
$25,000$ iterations burn in, 
using the adaptive random walk Metropolis algorithm with normal proposal described in \citet{Pham2014note}. 
The resulting estimated marginal posterior densities for $\lambda, \sigma, \xi$  are shown in
Figure~\ref{density_inclu}. The results for the proposed empirical likelihood based method (solid lines) agree quite well with rejection ABC (dotted lines) and synthetic likelihood (dashed lines). Similar to previous examples, however, there is a slight underestimation of posterior uncertainty in the empirical likelihood ABC method.  

The summary statistics in this example were judiciously chosen.  This dataset was also considered by \citet{Pham2014note}, who used $(L-112)/112$, the minimum, the maximum and the median of the observed inclusions as summaries in their analysis.
We realised that the observed summaries are too extreme for the summaries generated from the potentially mis-specified model for most values of $\theta$.  That is, for most values of the parameter, the problem in \eqref{eq:w2} was infeasible and the estimated empirical likelihood was zero.
As a result, the MCMC scheme to sample from the resulting posterior mixed very slowly.
However, the performance of the proposed method is no worse than the synthetic likelihood for these summaries.  It is well-known that (see \citet{price+dln16}), for these summaries the synthetic likelihood covariance matrix is often poorly estimated, resulting in gross over-estimation of the likelihood in the tail of the posterior, which leads to poor mixing in the MCMC algorithm.  
It turns out that (see \citet{frazier+rr17}), the simple rejection ABC is more robust for such potentially mis-specified models.

\section{Discussion}

We have developed a new and easy-to-use empirical likelihood ABC method.  For implementation, all that is required
are some summary statistics, their observed values, and the ability to simulate from the model.  Properties of the approach
have been explored both empirically and theoretically.  The method enjoys posterior consistency under reasonable conditions, 
and shows good performance in simulated and real examples with appropriate summary statistic choices.

The proposed method is based on an interpretable empirical likelihood.  Thus, unlike the conventional rejection ABC method, no tolerance or bandwidth needs to be specified.
  Furthermore, unlike the synthetic likelihood, the proposed method does not assume joint normaility of the summary statistics.  By using an argument similar to data cloning \citep{lele+dl07}, we avoid any requirement of parameter dependent constraints to determine the empirical likelihood as well.  
  This directly contrasts with the previous empirical likelihood based ABC methods \citep{mengersen+pr13}.

In the proposed method, the empirical likelihood is approximated using data simulated from the underlying generative model.  Empirical evidence suggests that, like the synthetic likelihood \citep{price+dln16}, it is not sensitive to the number of generated replications.
  On the other hand, if the joint normality of the summary statistics is not satisfied (like in the ARCH(1) model above), the proposed approach is seen to work better than the synthetic likelihood.
  Since no distributional assumptions are made, the proposed approach can avoid the additional burden of searching for suitable marginal transformations to improve adherence to such assumptions.  As a result, it can be easily automated in practical applications.

Even though simple choices often work for our method, a judicious selection of summary statistics is required.  As we have demonstrated above, summaries which fit the model rather poorly, may result in failures of the empirical likelihood based ABC.
However, for a poorly fitting model, such computational problems may arise for alternative methods as well.
It is important to diagnose poor model fit for the chosen summary statistics when this occurs (see \citet{frazier+rr17} for suggestions).
  Interestingly, synthetic likelihood can often down-weight unimportant summaries, which, as implemented, is not true for the proposed method. 
   Penalised empirical likelihood which can choose constraints has been recently considered. Such methods can be used in our proposed ABC as well. 

From the presented examples, it seems that the empirical likelihood slightly underestimates posterior uncertainty. Under-coverage of frequentist empirical likelihood confidence intervals is a well-known problem. This is most likely the Bayesian counterpart to that phenomenon.
The error would be small provided minimal and informative summary statistics are used.  Additionally, a wide variety of suggestions, similar to \citet{tsao2013,jing+tz17} etc. can be adapted in order to remedy this underestimation.  


Finally, similar to \citet{chernozhukov+h03}, it is likely that under suitable conditions, a Bernstein-von Mises theorem would hold for our posteriors, based on which asymptotic sandwich-type variance corrections might
also be considered.  We leave these investigations to future endeavours.  



\section*{Appendix}
\noindent{\it Proof of Lemma \ref{lem:1}.}  We show that for every $\epsilon>0$, there exists $n_0=n_0(\epsilon)$ such that for any $n\ge n_0$ for all $\theta\in\Theta_n$ the maximisation problem in \eqref{eq:w2} is feasible with probability larger than $1-\epsilon$.  

By assumption, for each $\theta$, random vectors $\xi^{(n)}_i(\theta)$ are i.i.d., put positive mass on each orthant and supremum of their lengths in each orthant diverge to infinity with $n$.  The random vectors $\left\{\xi^{(n)}_i(\theta)-\xi^{(n)}_o(\theta_o)\right\}$ will inherit the same properties.  
That is, there exists integer $n_0$, such that for each $n\ge n_0$, the convex hull of the vectors $\left\{\xi^{(n)}_i(\theta)-\xi^{(n)}_o(\theta_o)\right\}$, $i=1$, $\ldots$, $m(n)$, would contain the unit sphere with probability larger than $1-\epsilon/2$.  


We choose an $n\ge n_0$ and a $\theta\in\Theta_n$. For this choice of $\theta$:
\begin{align}
h^{(n)}_i(\theta,\theta_o)=&b_n\left\{\mathfrak{g}(\theta)-\mathfrak{g}(\theta_o)\right\}+\xi^{(n)}_i(\theta)-\xi^{(n)}_o(\theta_o)\nonumber\\
=&c_n(\theta)+\xi^{(n)}_i(\theta)-\xi^{(n)}_o(\theta_o),\nonumber
\end{align}
where, $\mid\mid\mathfrak{g}(\theta)-\mathfrak{g}(\theta_o)\mid\mid\le b^{-1}_n$.  That is, $\mid\mid c_n(\theta)\mid\mid\le 1$.  
Now, since $-c_n(\theta)$ is in the convex hull of the vectors $\left\{\xi^{(n)}_i(\theta)-\xi^{(n)}_o(\theta_o)\right\}$, $i=1$, $\ldots$, $m(n)$, with probability larger than $1-\epsilon/2$, there exists weights $w\in\Delta_{m(n)-1}$ such that,

\[
-c_n(\theta)=\sum^{m(n)}_{i=1}w_i\left\{\xi^{(n)}_i(\theta)-\xi^{(n)}_o(\theta_o)\right\}.
\]
Now it follows that for the above choice of $w$ that
\[
\sum^{m(n)}_{i=1}w_ih^{(n)}_i(\theta,\theta_o)=c_n(\theta)+\sum^{m(n)}_{i=1}w_i\left\{\xi^{(n)}_i(\theta)-\xi^{(n)}_o(\theta_o)\right\}=0,
\] 
which shows that the problem in \eqref{eq:w2} is feasible.\hfill $\square$

\bigskip


\noindent{\it Proof of Lemma \ref{lem:2}.} Let $\epsilon$ be as in the statement.  By assumption (A1), for some $\delta>0$,  $\mid\mid\mathfrak{g}(\theta)-\mathfrak{g}(\theta_o)\mid\mid>\delta$ for all $\theta$ with $\mid\mid\theta-\theta_o\mid\mid >\epsilon$.

Consider $\eta>0$. We show that there exists $n_0=n_0(\eta)$ such that for any $n\ge n_0$, the constrained maximisation problem in \eqref{eq:w2} is not feasible for all $\mid\mid\theta-\theta_o\mid\mid >\epsilon$, with probability larger than $1-\eta$. 


Let if possible $w\in\Delta_{m(n)-1}$ be a feasible solution.  Hence we get:

\begin{align}
0=&\sum^{m(n)}_{i=1}w_ih^{(n)}_i(\theta,\theta_o)=\sum^{m(n)}_{i=1}w_i\left\{g^{(n)}\left(X_i(\theta)\right)-g^{(n)}\left(X_o(\theta_o)\right)\right\}\nonumber\\
=&\left\{\mathfrak{g}^{(n)}(\theta)-\mathfrak{g}^{(n)}(\theta_o)\right\}+\left\{\sum^{m(n)}_{i=1}w_i\xi^{(n)}_i(\theta)\right\}-\xi^{(n)}_o(\theta_o),\nonumber
\end{align}
so that 
\begin{equation*}
-b_n\left\{\mathfrak{g}(\theta)-\mathfrak{g}(\theta_o)+o(1)\right\}=\sum^{m(n)}_{i=1}w_i\xi^{(n)}_i(\theta)-\xi^{(n)}_o(\theta_o).
\end{equation*}

By dividing both sides by $b_n$ we get:
\begin{equation}\label{eq:cons}
-\left\{\mathfrak{g}(\theta)-\mathfrak{g}(\theta_o)\right\}=\sum^{m(n)}_{i=1}w_i\left\{\frac{\xi^{(n)}_i(\theta)}{b_n}-\frac{\xi^{(n)}_o(\theta_o)}{b_n}\right\}-o(1).
\end{equation}
Now, $\mid\mid\xi^{(n)}_o(\theta_o)\mid\mid/b_n\le \sup_{i\in\{o,1,2\ldots,m(n)\}}\mid\mid\xi^{(n)}_o(\theta_o)\mid\mid/b_n$ and
\begin{align}
\left|\left|\sum^{m(n)}_{i=1}w_i\frac{\xi^{(n)}_i(\theta)}{b_n}\right|\right|\le&\sum^{m(n)}_{i=1}w_i\frac{\mid\mid\xi^{(n)}_i(\theta)\mid\mid}{b_n}\nonumber\\
\le&\sup_{i\in\{o,1,2\ldots,m(n)\}}\frac{\mid\mid\xi^{(n)}_i(\theta)\mid\mid}{b_n}.\nonumber
\end{align}
That is, by assumption (A3), there exists $n_0(\eta)$ such that for any $n\ge n_0$, the RHS of \eqref{eq:cons} is less than $\delta$ for all $\theta\in B(\theta_o,\epsilon)$, with probability larger than $1-\eta$.  However, $\mid\mid\mathfrak{g}(\theta)-\mathfrak{g}(\theta_o)\mid\mid>\delta$. 
We arrive at a contradiction.  Thus the problem is infeasible for every $\theta\in B(\theta_o,\epsilon)^C$ with probability larger than $1-\eta$.\hfill $\square$




\bigskip

\noindent{\it Proof of Theorem \ref{thm:1}.}  Let $s(\theta)$ be a continuous, bounded function. We choose an $\epsilon>0$. Then by Lemma \ref{lem:1}, there exists $n(\epsilon)$, such that for any $n>n(\epsilon)$ and $\theta\in B\left(\theta_o,\epsilon\right)^C$, 
the posterior $\Pi_n\left(\theta\mid X_o(\theta_o)\right)=0$.  That is for any $n>n(\epsilon)$, 
\begin{align}
~&\int_{\Theta}s(\theta)\Pi_n\left(\theta\mid X_o(\theta_o)\right)d\theta=\int_{B\left(\theta_o,\epsilon\right)}s(\theta)\Pi_n\left(\theta\mid X_o(\theta_o)\right)d\theta\nonumber\\
=&\int_{B\left(\theta_o,\epsilon\right)}\left\{s(\theta)-s(\theta_o)\right\}\Pi_n\left(\theta\mid X_o(\theta_o)\right)d\theta\nonumber\\
&+s(\theta_o)\int_{B\left(\theta_o,\epsilon\right)}\Pi_n\left(\theta\mid X_o(\theta_o)\right)d\theta.\nonumber
\end{align}

Since the function $s(\theta)$ is bounded and continuous at $\theta_o$, the first term is negligible.  Furthermore, $\int_{B\left(\theta_o,\epsilon\right)}\Pi_n\left(\theta\mid X_o(\theta_o)\right)d\theta=1$.  This implies the integral converges to $s(\theta_o)$.  This shows, the posterior converges weakly to $\delta_{\theta_o}$.  \hfill $\square$





\section*{Acknowledgments}
Sanjay Chaudhuri was supported by a Singapore Ministry of Education Academic Research Fund Tier 1 grant (R-155-000-176-114).  Subhroshekhar Ghosh was supported by National University of Singapore grant R-146-000-250-133.
David Nott was supported by a Singapore Ministry of Education Academic Research
Fund Tier 1 grant (R-155-000-189-114).
  Pham Kim Cuc was supported by the Singapore-Peking-Oxford Research Enterprise, COY-15-EWI-RCFSA/N197-1.



\bibliographystyle{chicago}


\end{document}